\documentclass[10pt,conference]{IEEEtran}
\IEEEoverridecommandlockouts
\usepackage{cite}
\usepackage{url}

\usepackage{titletoc}
\usepackage{amsmath,amssymb,amsfonts}
\usepackage{algorithmic}
\usepackage{graphicx}
\usepackage{textcomp}
\usepackage{bm}
\usepackage{booktabs}
\usepackage{threeparttable}
\usepackage{multirow}
\usepackage{tikz}
\usepackage{pgfplots}
\def\BibTeX{{\rm B\kern-.05em{\sc i\kern-.025em b}\kern-.08em
    T\kern-.1667em\lower.7ex\hbox{E}\kern-.125emX}}
\pgfplotsset{compat=1.14}

\usepackage[utf8]{inputenc}
\usepackage{listings}
\usepackage{color}

\newcommand{\lijin}[1]{{\color{blue} Lijin: #1}}
\newcommand{\toolname}{EVulHunter}
\newcommand{\vula}{\textit{fake EOS transfer}}
\newcommand{\vulb}{\textit{fake transfer notice}}

\begin{document}

\title{EVulHunter: Detecting Fake Transfer Vulnerabilities for EOSIO's Smart Contracts at Webassembly-level 
}

\author{
\IEEEauthorblockN{Lijin Quan$^{1}$, Lei Wu$^{2}$, Haoyu Wang$^{1}$}
\IEEEauthorblockA{
$^{1}$ Beijing University of Posts and Telecommunications,
$^{2}$ PeckShield, Inc.}
}

\maketitle

\begin{abstract}
As one of the representative Delegated Proof-of-Stake (DPoS) blockchain platforms, EOSIO's ecosystem grows rapidly in recent years. A number of vulnerabilities and corresponding attacks of EOSIO's smart contracts have been discovered and observed in the wild, which caused a large amount of financial damages. However, the majority of EOSIO's smart contracts are not open-sourced. As a result, the WebAssembly code may become the only available object to be analyzed in most cases. Unfortunately, current tools are web-application oriented and cannot be applied to EOSIO WebAssembly code directly, which makes it more difficult to detect vulnerabilities from those smart contracts.
In this paper, we propose \toolname, a static analysis tool that can be used to detect vulnerabilities from EOSIO WASM code automatically. We focus on one particular type of vulnerabilities named \textit{fake-transfer}, and the exploitation of such vulnerabilities has led to millions of dollars in damages. To the best of our knowledge, it is the first attempt to build an automatic tool to detect vulnerabilities of EOSIO's smart contracts. The experimental results demonstrate that our tool is able to detect fake transfer vulnerabilities quickly and precisely.
EVulHunter is available on GitHub\footnote{Tool and benchmarks: https://github.com/EVulHunter/EVulHunter}
and YouTube\footnote{Demo video: https://youtu.be/5SJ0ZJKVZvw}.
\end{abstract}

\section{Introduction}
\label{sec-intro}
With the growing prosperity of cryptocurrencies like Bitcoin, blockchain techniques has become more attractive and been adopted by different aeras, such as financial and logistical systems. Due to the limited throughput (e.g., Transaction Per Second, aka TPS) derived from the essence of the Proof-of-Work (PoW) consensus, traditional blockchain platforms (e.g., Bitcoin and Ethereum),
cannot be used to support high-performing applications. Researchers proposed different consensuses, such as Proof-of-Share (PoS) and Delegated Proof-of-Stake (DPoS), to resolve the performance issue. 

As one of the most representative DPoS platforms and the first decentralized operating system, EOSIO~\cite{eos} has become one of the most active communities all over the world. 
EOSIO adopts a multi-threaded mechanism based on its DPoS consensus protocol~\cite{eos-wiki}. EOSIO claims that it is capable of achieving millions of TPS. The main native token of EOSIO is called \textit{EOS}. The amount of EOS tokens of the holders are used to allocate system resources (e.g., bandwidth and storage), vote and participate in the on-chain governance, according to the corresponding proportion of the total stake. Briefly speaking, 21 Block Producers (BPs) are voted during its launch, and a validate block is generated every 500 ms block time by those BPs in a round-robin schedule. 

Smart contract is a computer protocol which allows users to digitally negotiate an agreement in a convenient and secure way~\cite{ethereum-wiki}. In contrast to the traditional contract law, the transaction costs of the smart contract are dramatically reduced, and the correctness of its execution is ensured by the consensus protocol. EOSIO's smart contracts can be written in C++
, which will be compiled down to WebAssembly (aka WASM) and executed in EOSIO WASM virtual machine. WASM is a web standards specifying the binary instruction format for a stack-based virtual machine, and it can be run in modern web browsers and other environments~\cite{wasm-wiki}.

However, it is not easy to guarantee the security of the implementation of smart contracts, new platforms in particular. A number of vulnerabilities have been discovered from EOSIO's smart contracts, including fake EOS transfer, fake transfer notice and flawed random numbers generators. As such, severe attacks have been observed in the wild, which caused a large amount of financial damages. 

Unfortunately, most Decentralized Applications (DApps) projects on EOSIO are not open-sourced, while the web-application oriented analyzed tools cannot be applied to EOSIO WASM code directly, which makes it more difficult to detect vulnerabilities from those smart contracts. Specifically, there does exist two major challenges must be overcome. First of all, since the EOSIO WASM code does not contain context information (e.g., symbols, function names and type information), we have to infer the indexes of the functions we are interested in, including relevant system functions and functions implemented by the developer. Secondly, the Contract Development Toolkit (CDT) of EOSIO has been evolving ever since the launch of the mainnet, and we have to cover variants because of the changes caused by CDT(s).

In this paper, we propose a static analysis tool named \textit{\toolname} which can be used to detect vulnerabilities from EOSIO WASM code automatically. Specifically, it first traverses the WASM code and constructs the corresponding Control Flow Graphs (CFGs), and then detects the existence of vulnerabilities based on predefined patterns. To facilitate the deep analyses, {\toolname} also contains a WASM Simulator module which is capable of performing the simulated code execution. Besides, we focus on one particular type of vulnerabilities named \textit{fake-transfer} with two variants, ~\textit{fake EOS transfer} and \textit{fake transfer notice}. The exploitation of fake-transfer vulnerability has led to millions of dollars in damages from gambling DApps.

To the best of our knowledge, our work is the first attempt to build an automatic tool to detect vulnerabilities of EOSIO's smart contracts. The experimental results demonstrate that our system is able to detect target vulnerabilities quickly and precisely with low false positives. Our efforts can positively shed some light on the exploration of vulnerability detection of EOSIO's smart contracts.
\section{Background}

\subsection{EOSIO's Smart Contract}
In EOSIO, an \textit{action} (a base32 encoded 64-bit integer) is used to represent a single operation, which can be sent individually, or in a combined form, to serve the purpose of communication between a smart contact and an account~\cite{eos-dev}. A \textit{transaction} is composed of one or more actions. 

In order to handle requested actions, an \textit{apply} function is necessary to dispatch corresponding action handlers for any validate smart contract. Specifically, \textit{apply} function listens to all incoming actions and execute the concrete action handlers accordingly. To this end, the input parameters of \textit{apply} function, i.e., the \textit{receiver} (the account that is currently processing the action), \textit{code} (the account that authorized the contract), and \textit{action} (the ID of the currently running action), are used as filters to map to the desired functions that implement particular actions~\cite{eos-dev}.

As one of the 5 fundamental smart contracts of EOSIO\footnote{Namely EOSIO.contracts, i.e., \textit{eosio.bios}, \textit{eosio.token}, \textit{exchange}, \textit{eosio.msig}, \textit{eosio.system}}, \textit{eosio.token} contract is a token standard (like Ethereum's ERC20 standard) contract which enables the creation of many different tokens all running on the same contract but potentially managed by different users~\cite{eos-doc}. It can be used to \textit{create}, \textit{issue} and \textit{transfer} tokens~\cite{eos-dev}, and the \textit{transfer} functionality is related to the target \textit{fake-transfer} vulnerabilities. 


\subsection{Fake-Transfer Vulnerabilities}
\begin{figure}[htbp!]
    \centering
    \includegraphics[width=0.5\textwidth]{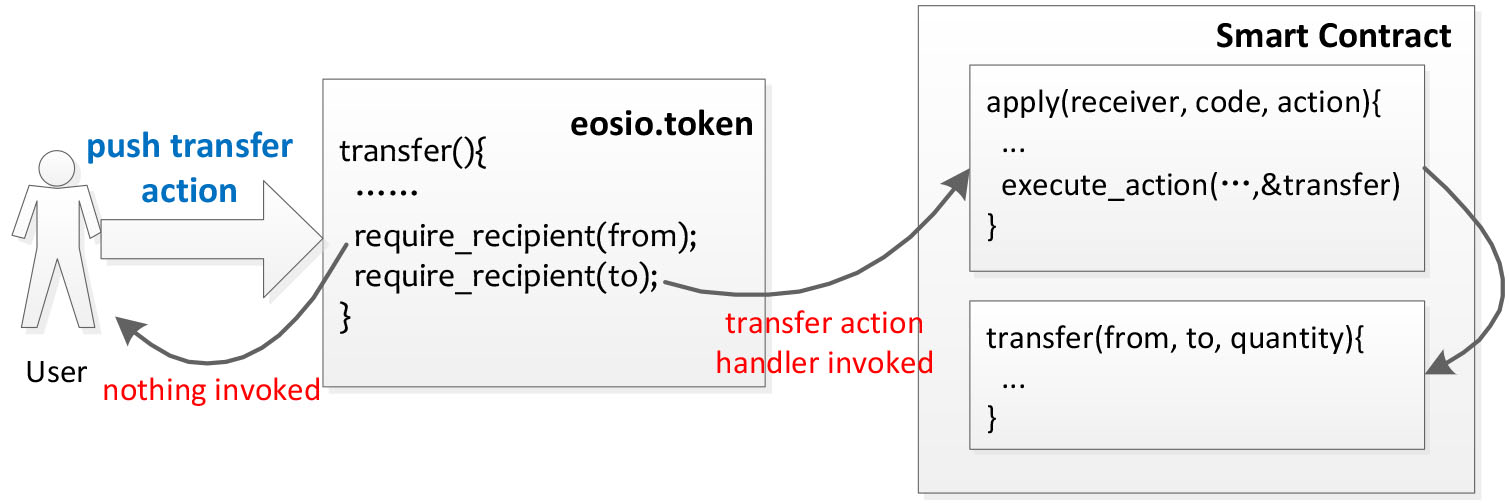}
    \caption{The lifecycle of a transfer action.}
    \label{fig-transfer-lifecycle}
\end{figure}

Figure~\ref{fig-transfer-lifecycle} shows the lifecycle of a \textit{transfer} action. Firstly a user sends the \textit{transfer} action to the \textit{eosio.token} contract. After that, the \textit{eosio.token} contract will modify the balances accordingly, and then launch the following two invocations: \textit{\lstinline{require_recipient (from)}} and \textit{\lstinline{require_recipient (to)}}, \textit{from} and \textit{to} are source account and destination account respectively. Finally, if any account has been deployed with a contract, and there does exist a pre-defined \textit{transfer} action handler in this contract, the handler would be invoked.

\noindent \textbf{Fake EOS Transfer.}
As mentioned earlier, the \textit{code} parameter of the \textit{apply} function represents the account that authorized the contract. Obviously, the \textit{code} parameter of the \textit{apply} function of the victim (i.e., the recipient \textit{to}) should be \textit{eosio.token} in a normal \textit{transfer} action. However, if an implementation of the \textit{apply} function did not verify the \textit{code} parameter properly, it might be deceived into receiving fake EOS tokens and then executing the further code logic.

\noindent \textbf{Fake EOS Notice.}
It is an advanced variant of \textit{fake-transfer} vulnerabilities with a well-implemented \textit{apply} function. However, two parameters of the \textit{transfer} function, i.e., \textit{to} and \textit{\_self}, have not been checked for equality. Specifically, the attacker was able to create an intermediate contract that forwarding the incoming \textit{transfer} action to the victim contract. As such, the victim contract would be misled into believing it was receiving EOS tokens.


\section{Approach}

In this section, we will first introduce the overall design of the proposed \textit{\toolname} system, and then depict each component individually. 

\begin{figure}[h!]
    \centering
    \includegraphics[width=0.5\textwidth]{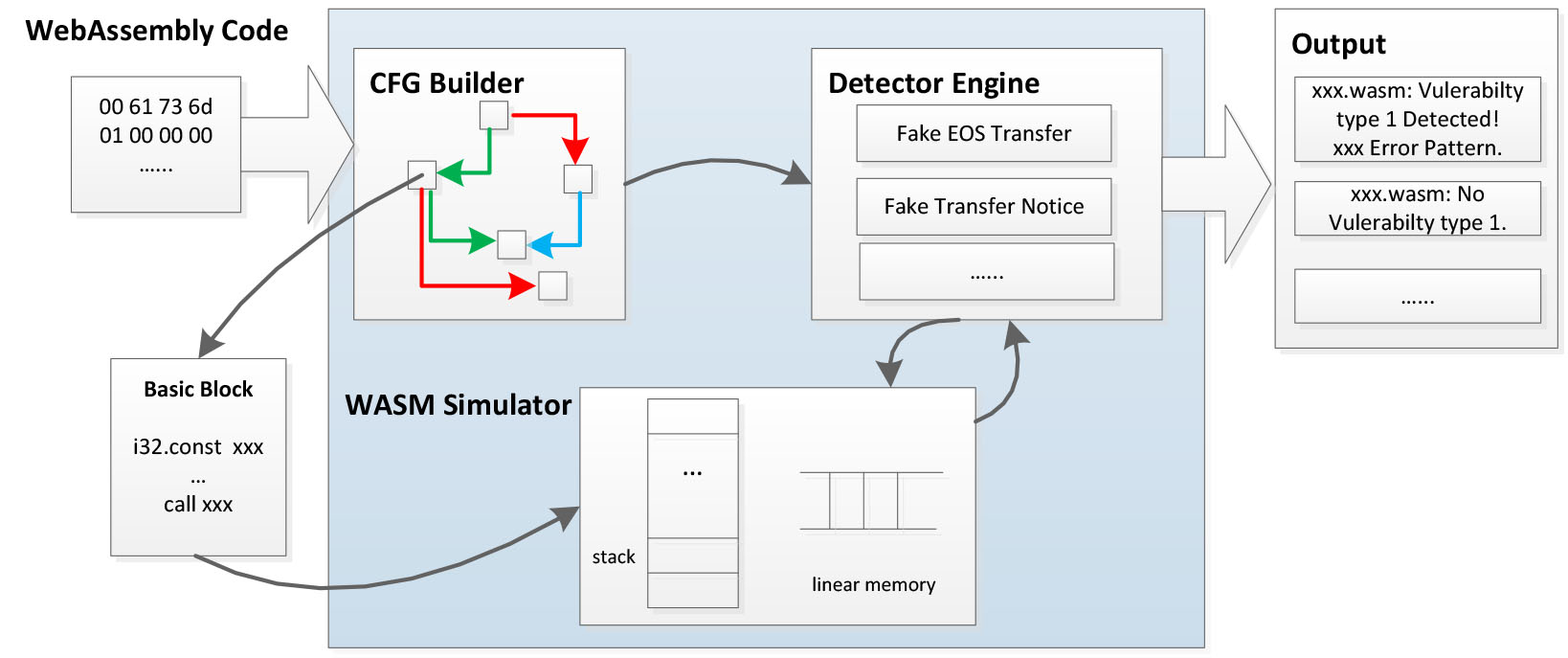}
    \caption{Overview of \toolname.}
    \vspace{-0.1in}
    \label{fig:arch}
\end{figure}

\subsection{Overview}
Figure~\ref{fig:arch} illustrates the framework of the proposed system, which accepts the EOSIO WASM code as the input. EVulHunter is composed of three modules, including \textit{CFG Builder}, \textit{WASM Simulator} and \textit{Detector Engine}.  

We first build the CFGs based on \textit{Octopus}\footnote{https://github.com/quoscient/octopus}, an open source project that is able to parse EOSIO WASM code. Specifically, the CFGs of the \textit{apply} function will be generated in our CFG builder. After that, the entire CFGs will go through the Detector Engine module. To facilitate the analyses, specific Basic Blocks (BBs) might be selected to be executed by the WASM Simulator on demand, where the stack the linear memory would be rebuilt to serve the purpose. 

We will describe \textit{WASM Simulator} and \textit{Detector Engine} in the following subsections, \ref{sub-sec:wasm} and \ref{sub-sec:detector} respectively.



        
        
        


\subsection{WASM Simulator}
\label{sub-sec:wasm}
\textit{WASM Simulator} is designed to be a simple, concise and generic  Virtual Machine (VM) to support further analyses with high extensibility. Basically, this VM keeps a \textit{Stack} and \textit{Memory} structure, which will be modified during tracing instructions of the WASM code. To facilitate the analysis, several special patterns have been observed and summarized, including \textit{\_self}, \textit{from}, \textit{to} and strings in a format of 32-bit encoding integer, so that the simulator is able to recover the semantic type information by mimicking the code execution. For example, \textit{get\_local 7} means a popped value from the stack will be assigned to a variable named \textit{local7}, even without the knowledge of the value type. After simulated execution, it is possible to determine that \textit{local7} represents a base32 encoded 64-bit integer and is exactly the ``\textit{eosio.token}'' in the context.

\noindent \textbf{Locating Indirect Call Functions.}
The WASM Simulator can be used to locate indirect call functions in the \textit{apply} function, which is the first challenge mentioned in Section~\ref{sec-intro}. 
Specifically, the link between the \textit{apply} function and the target function is implemented with the indirect call mechanism by using an API named \lstinline{execute_action()} in all the versions of CDT. Although different CDT may have different implementations, the invariant feature is that the indirect call function pointer is always the last parameter of the  \lstinline{execute_action()} function. As the top item in the stack is always the index number of the indirect call function when calling \lstinline{execute_action}, we can use this feature to determine the indexes of the functions in the corresponding WASM code.



\subsection{Detector Engine}
\label{sub-sec:detector}

All detectors are wrapped in the \textit{Detector Engine} module, which may conduct deep analyses by interacting with the WASM Simulator when necessary.
We have implemented two detectors for {\vula} and {\vulb} vulnerabilities respectively, as follows:

\begin{itemize}\label{assumption:vul1}
\item In case of {\vula}, if none of the functions pre-defined by the developer is likely to be invoked when the incoming action is not authorized by {\textit{eosio.token}}, which further suggests that the {\textit{transfer}} function won't be invoked as well, therefore it is not vulnerable to {\vula} vulnerability; otherwise it is tagged as a vulnerable contract.
    
\item In case of {\vulb}, if there exists a comparison between {\textit{to}} and \textit{\_self} in the \textit{transfer} function, it is not vulnerable to {\vulb} vulnerability; otherwise it is tagged as a vulnerable contract.
\end{itemize}

\begin{figure}
    \centering
    \includegraphics[width=0.5\textwidth]{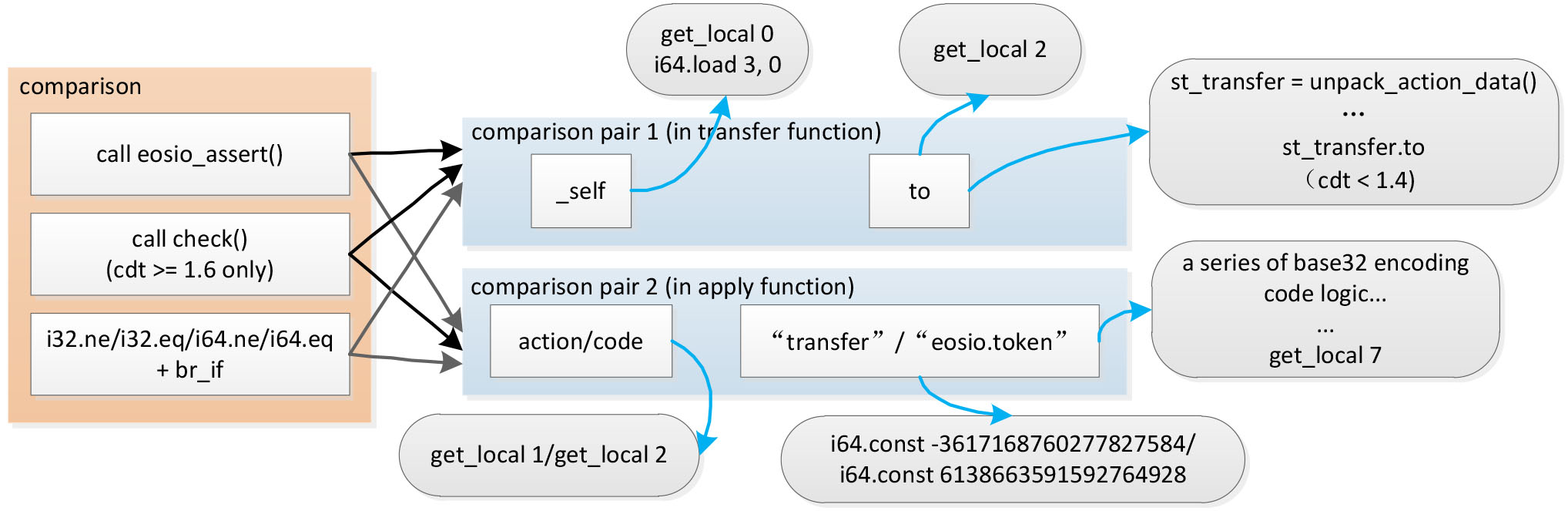}
    \caption{Summary of Covering Variants.}
    \label{fig:branches}
\end{figure}

\noindent \textbf{Covering Variants.}
We have to handle two types of variants. Firstly, variants derived from the the evolutions of CDT. Secondly, variants come from one specific verion of CDT, by which even the same logic might be served by different groups of WASM instructions. We have tried our best to summarize and cover cases as many as possible, if not all, to make the detectors robust and complete. 
All variants being analyzed are summarized in Fig.~\ref{fig:branches}, including comparison patterns, comparison pairs and corresponding descriptions of elements being compared shown in elliptical boxes. There are 3 comparison patterns and 2 comparison pairs respectively, implying 6 combinations in total.

\section{Evaluation}

\begin{table}[t]
\caption{The Distribution of our benchmark.}
\label{table:benchmark}
\centering

\begin{tabular}{l|l|l|l}
\hline
                  & \# Vulnerable & \# Non-vulnerable & \# Total \\ \hline
Fake EOS Transfer &    75           &        109    &     184     \\ \hline
Fake EOS Notice   &   141            &        54    &   195       \\ \hline \hline
Total            &   159            &     82       &    241      \\ \hline
\end{tabular}
\end{table}

\begin{table*}[t]
\caption{The Overall Experiment Result.}
\label{table:result}
\centering

\begin{tabular}{l|c|c|c|c|c|c|c}
\hline
                  & True Positive & False Positive & True Negative & False Negative & Precision & Recall & Accuracy \\ \hline
Fake EOS Transfer &     75          &      26          &       83       &         0       &     74.26\%  & 100\% & 85.87\%    \\ \hline
Fake EOS Notice   &       141       &  0              &        54      &      0         &    100\% & 100\% &    100\%  \\ \hline
Total             &        216       &     26           &       137        &     0          & 89.26\%    & 100\% & 93.14\% \\  \hline
\end{tabular}
\end{table*}

\subsection{Benchmark}
It appear that no available benchmarks on vulnerable EOS smart contracts in our community could be used for evaluation. Thus, we propose to manually craft a benchmark from the known reported vulnerable Dapps\footnote{https://github.com/peckshield/EOS/tree/master/known\_dapp\_attacks}. For example, it is reported that EOSBet contract was attacked by exploiting the fake EOS transfer vulnerability in September 2018, and it suffered from the fake EOS notice vulnerability in October 2018. In this way, we seek to identify the corresponding vulnerable versions of the reported smart contracts, and then collect the patched ones with no vulnerabilities.

At last, we have collected 241 EOS smart contracts in total (with 159 vulnerable ones), 184 of them were used as the benchmark of fake EOS transfer vulnerability, and 195 of them were used to evaluate the fake EOS notice vulnerability detection. The distribution of our benchmark is shown in Table~\ref{table:benchmark}. Note that one smart contract could have both fake EOS transfer and fake EOS notice vulnerabilities.

\subsection{Detection Result}

\noindent \textbf{Overall Result.} Table~\ref{table:result} shows the overall detection result. 
For the fake EOS transfer vulnerability, our tool could achieve an overall accuracy of 86\%, while for the fake EOS notice vulnerability, we could achieve an overall accuracy of 100\%. 
This result suggests that EVulHunter is able to detect the fake notice vulnerabilities with high precision and recall.
No false negatives were found in our evaluation, while 26 false positives were found when detecting fake EOS transfer vulnerability.

\noindent \textbf{False Positives.}
We have conducted a manual investigation for all the 26 false positives, which were finally attributed to the history versions of contract ``eosbetdice11''.
Specifically, besides the account \textit{eosio.token}, these contracts acknowledge the legality of an extra account named \textit{eosbettokens}~\footnote{Probably related to tokens they issued to support their own business.}, which breaks our rules introduced in \ref{assumption:vul1}. 

In EOSIO platform, any account can deploy a token contract, however, it is not trivial for us to collect all possible legal accounts in an anonymous audit. To alleviate this issue, our system could be easily extended to support a whitelist mechanism which allows developers to customize extra legal accounts by themselves.

\subsection{Performance}

We further evaluate the performance of \toolname, e.g., the time cost to detect fake transfer vulnerabilities in a given smart contracts.
For the 242 smart contracts in our benchmark, we have recorded the detection time of each contract, and then analyzed the correlation between the size of smart contract and the detection time.

\begin{figure}[t]
    \centering
    \includegraphics[width=0.45\textwidth]{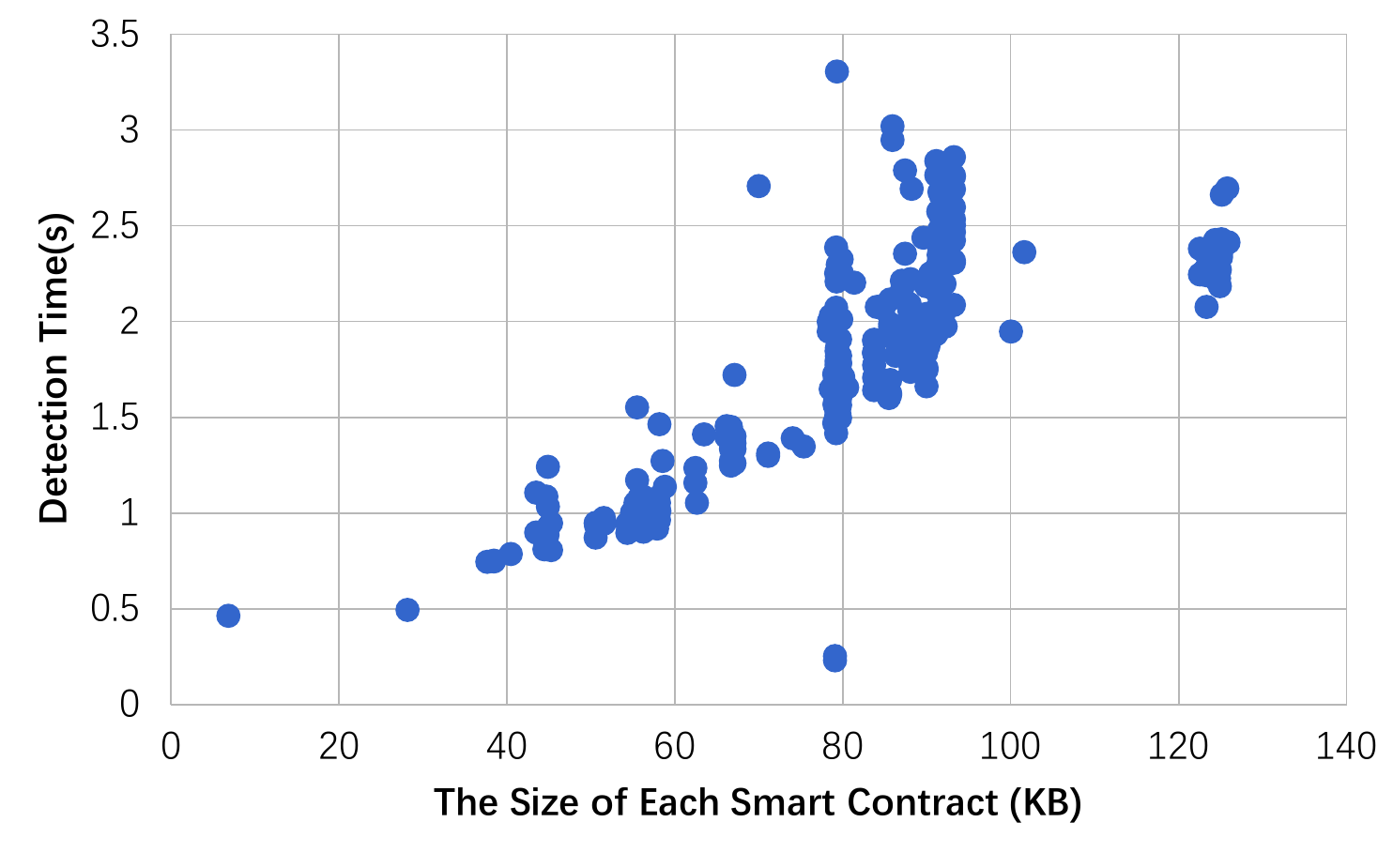}
    \vspace{-0.1in}
    \caption{The Performance Evaluation of \toolname.}
        \vspace{-0.1in}
    \label{fig:performance}
\end{figure}

As shown in Figure~\ref{fig:performance}, it is obvious that time consumption of vulnerability detection increased linearly with the increasing of the size of smart contract, although the increasing rate is very small.
In general, the detection time rages from 1 second to 3 seconds.
This result suggests that EVulHunter is efficient to flag the vulnerabilities in EOS smart contracts, which could be easily scale to thousands of smart contracts.

\section{Related Work}

The security vulnerabilities of smart contracts have attracted great attention of our research community. However, almost all of the previous studies focus on analyzing the vulnerabilities in the Ethereum smart contracts~\cite{liu2018reguard, atzei2017survey, luu2016making, tikhomirov2018smartcheck, jiang2018contractfuzzer, tsankov2018securify}, as a large number of smart contracts in the Ethereum ecosystem are open-sourced, and a number of analyzing tools could be leveraged.
For example, Liu \textit{et al.}~\cite{liu2018reguard} presented an analyzer aimed at reentrancy, the type of vulnerability which cause `TheDAO' hard fork on Ethereum. Atzei \textit{et al.}~\cite{atzei2017survey} covered more security vulnerabilities and showed a series of attack which allow attacker to steal money from smart contract. Luu \textit{et al.}~\cite{luu2016making} built a sysbolic execution tool to find potential security bugs. Tikhomirov \textit{et al.}~\cite{tikhomirov2018smartcheck} provided a static analysis tool to detect problematic language constructs.
Our work is the first attempt to detect fake transfer vulnerabilities for EOSIO's smart contracts. We believe our efforts could shed some light on the detection of vulnerabilities in the EOSIO's smart contracts.

\section{Conclusion}
In this work, we present the first systematic attempt to automatically detect \textit{fake-transfer} vulnerabilities from EOSIO WASM code. Specifically, it first traverses the WASM code and constructs the corresponding Control Flow Graphs (CFGs), and then detects the existence of vulnerabilities based on predefined patterns. The evaluation results demonstrate the system performance, including speed and accuracy. 

There are a number of future lines of
work we will explore. For example, our detection simply relies on constant patterns, which may not be suitable to model the characteristics of other types of vulnerabilities. As such, some advanced program analysis techniques (e.g., symbolic execution) could be used to achieve more accurate results. To this end, we may have to revise the design of our system to support advanced analyses. Despite the limitations, we still believe our efforts and observations could positively contribute to the community.


\end{document}